\RequirePackage{fix-cm}
\documentclass[11pt]{article}       % onecolumn (second format)

\usepackage{graphicx}

\usepackage[T1]{fontenc}

\usepackage{mathptmx}      % use Times fonts if available on your TeX system
\usepackage{amsfonts}

\newtheorem{peqm}{Postulate EQM}

\begin{document}
\fontfamily{ptm}\selectfont
\title{Everett's Missing Postulate and the Born Rule%\thanks{Grants or other notes
}
%\subtitle{}

%\titlerunning{Everett's Missing Postulate and the Born Rule}        % if too long for running head

\author{Per Arve}


%\institute{Per Arve \at
%              Johan Enbergs väg 19 \\
%              17162 Solna, Sweden \\
%              Tel.: +46-70-2496694\\
%              ORCID ID 0000-0002-1432-3316\\
%              \email{per.arve@me.com}           %  \\             
%}


\maketitle

\begin{abstract}
Everett's Relative State Interpretation has gained increasing interest due to the progress of understanding the role of decoherence. In order to fulfill its promise as a realistic description of the physical world, two postulates are formulated. In short they are 1) for a system with continuous coordinates $\vec{x}$, discrete variable $j$, and state $\psi_j(\vec{x})$, the density $\rho_j(\vec{x})=|\psi_j(\vec{x})|^2$ gives the distribution of the location of the system with the respect to the variables $\vec{x}$ and $j$; 2) an equation of motion for the state $i\hbar \partial_t \psi = H\psi$. The first postulate connects the mathematical description to the physical reality, which has been missing in previous versions. The contents of the standard (Copenhagen) postulates are derived, including the appearance of Hilbert space and the Born rule. The approach to probabilities earlier proposed by Greaves replaces the classical probability concept in the Born rule. The new quantum probability concept, earlier advocated by Deutsch and Wallace, is void of the requirement of uncertainty.  

%\keywords{Everett's Interpretation \and Born Rule \and Probability}
%\PACS{03.65.Ta,Ca,Ud,Yz}
\end{abstract}

\section{Introduction}

  In 1957 Everett \cite{everett1957relative} analyzed the quantum measurement process in terms of the unitary evolution of a total quantum state that describe not only the system to be measured, but also the apparatus, the observers of the apparatus and the observers of the observers.\footnote{Relational Quantum Mechanics \cite{sep-qm-relational,rovelli1996relational} is in some respects similar but differs in that observers should not describe themselves in terms of a `universal' wavefunction.}
  Several shortcomings initially plagued the interpretation that Everett introduced. The solution to the `preferred basis' problem seems to be solved by the decoherence theory. Another question that was not sufficiently well addressed was the Born rule. That this problem remained rather recently is clear from the discussions by Weinberg \cite{weinberg2015lectures} and the criticism by Hemmo and Pitowsky \cite{hemmo2007quantum}. New attempted proofs are still put forward. Wallace \cite{wallace2010prove,wallace2012emergent} has made a very elaborate proof that improves on Deutsch \cite{deutsch1999quantum} use of decision theory. The proofs by Sebens and Carroll \cite{carroll2014many,sebens2016self} \cite{pittphilsci14389} utilize the kind of `self-locating' uncertainty introduced by Vaidman \cite{pittphilsci8558} have been criticized by Kent \cite{kent2015does}. The critique by Kent is concerned with the lack of justification for the use of classical probability\footnote{Classical probability refers to situations in which there is uncertainty about a definite outcome.} in a situation which is inherently quantal. This critique may be equally appropriate against the recent attempt by McQueen and Vaidman \cite{pittphilsci14389}\footnote{They consider an agent that is located in different identical rooms but has not yet observed the result. The agent’s mind has not yet split; thus, it is not a classical uncertainty situation.}.

Another criticism from Kent \cite{KentAgainstMWI} and Maudlin \cite{maudlin2014critical} is directed to the lack of clear statements that define the theory. As Everett's quantum mechanics has abandoned the traditional postulates, there is a need for a new set of postulates. Everett assumed that the quantum state belongs to a Hilbert space, but this is a somewhat abstract mathematical notion. Everett took the position that ``The wave function is taken as the basic physical entity with {\em no a priori interpretation}. Interpretation comes after an investigation of the logical structure of the theory.''  However, if we are to understand how the notion of probability appears, we need a firm grip on the interpretation of the quantum state. This article offers the interpretation that Everett postponed.

The problematic relation between the theory of decoherence and any derivation of the Born rule emphasizes the need for a better starting point of the theory. The decoherence theory is based on the Born rule, but the decoherence theory is needed in order to give the branches for which the Born rule should give the probabilities. This relation constitutes an unacceptable circularity \cite{joos2000elements}. 

Wallace has relied in discussions of decoherence on that the Hilbert norm is a proper measure of what is important or negligible, thus avoid the reliance on the Born rule. Wallace argues, ``the Hilbert space norm is a perfectly objective feature of the physics, before any considerations of probability.'' However, that raises the question: Which are the reasons for the use of the Hilbert space, and its norm? 

Some attempts to prove the Born rule rely on the assumption, but without proof, that probabilities are conserved under unitary transformations \cite{carroll2014many,sebens2016self,wallace2010prove,BornRuleEnvariance,QDarwinPhysToday}, that probability is local quantity in space, or that probability is independent of what happens later \cite{everett1957relative,wallace2010prove}. Several approaches to the Born rule are only addressing a situation after the measurement \cite{carroll2014many,pittphilsci14389,sebens2016self} or after a pre-measurement \cite{BornRuleEnvariance,QDarwinPhysToday}. These shortcomings calls for a better derivation of the Born rule in the context of Everett's quantum mechanics.

For the analysis to be convincing, it is pivotal to have a definite and physically motivated starting point, that everything is derived from clearly stated assumptions \cite{barrett2017typical}, and the discussion of the Born rule probabilities address the situation prior to measurement. Not even the applicability of the probability concept can be taken for granted, as is illustrated by the criticism from Albert \cite{albert2010probability} and Kent \cite{kent2010one}.

\section{Postulates \label{postulates}}

If measurements are to be described by Everett's Quantum Mechanics (EQM), a wave packet entering the volume of a detector has to correspond to that the particle enters the detector. To conclude this, we have to rely on something other than the intuition that we have gained from using the standard (Copenhagen) postulates as they are abandoned in EQM.
To this end, two postulates are formulated. The first postulate address how the wavefunction describes position; the second defines the dynamics.
\begin{peqm}
\label{post:State}
The quantum state: The state is a set of complex functions of positions 
\begin{equation}
  \Psi =\{\psi_{jk}(t,\vec{x}_1,\vec{x}_2, ...)\}
  \label{Psi}
\end{equation}
where index $k$ is for gauge components and $j$ is a composite index for the spin components of all particles. Its basic interpretation is given by that the density 
\begin{equation}
\rho_j(t, \vec{x}_1,\vec{x}_2,\ldots) = \sum_k |\psi_{jk}(t, \vec{x}_1,\vec{x}_2, \ldots)|^2
\label{density} 
\end{equation}
answers where the system\footnote{The system can be a reasonably isolated system or the whole universe. The latter needs a quantum gravity formulation to be adequately addressed, which is beyond the present study.} is in position and spin. It is absolute square integrable and normalized to one
\begin{equation}
\int\! \int \!\cdots dx_1dx_2 \cdots \sum_{jk} |\psi_{jk}(t, \vec{x}_1,\vec{x}_2, \ldots)|^2 = 1.
\label{norm}
\end{equation}
This requirement signifies that the system has to be somewhere, not everywhere. 
If the value of the integral is zero, the system does not exist anywhere.
\end{peqm}

With the usual way of writing the norm $ \| \cdot \| $, equation (\ref{norm}) can be written $\| \Psi \|^2 = 1$.
If something is measurable, then it is possible to separate such a small part from the rest\footnote{The separation is here meant to be the process of preparation of experiments with all the complications that such a process entails. For example, the preparation of beam particles and targets in collision experiments.}. The separated part will act as a system of its own, thus cannot have zero norm. 
The difference between two states $\Psi$ and $\Psi'$ for which $\| \Psi - \Psi' \| = 0$ can have no measurable consequences, as $ \Psi - \Psi'$ is, according to EQM 1, physically equivalent to a function which is zero everywhere. 
This equivalence implies that the state of the system can be viewed as a vector in the Hilbert space of functions of the type (\ref{Psi}), the $L^2$ Hilbert space. 

The state vector $\psi$ is not directly observable as it is gauge dependent, while the density (\ref{density}) is independent of gauge and is, in principle, an observable quantity. The density, the distribution for where the system is located, gives how much the system is present at a location in configuration space $\vec{x}_1, \vec{x}_2, \ldots $, with the discrete index $j$. The density can be denoted the position distribution or the {\em presence} distribution, and both will be used here. The quantity {\em presence} has previously been denoted {\em measure of existence} by Vaidman \cite{vaidman1998schizophrenic} and {\em caring measure} by Greaves \cite{greaves2004understanding,greaves2007everettian}, but they have not fully clarified its meaning. The context in which they discussed the meaning of the quantity $\rho$ was that of probabilities and the Born rule. They did not derive the Born rule from their concepts, but the EQM 1 turns out to be a powerful starting point to prove the Born rule.

In EQM1, there is no mention of any relation between the density (presence) (\ref{density}) and probability. When the propagation of different parts is dependent on each other due to coherence, the concept of probability is not relevant. 
But, the density $\rho_j(t, \vec{x}_1, \vec{x}_2,\ldots)$ as a distribution of the particles positions is always relevant. It is similar to Schr\"odinger's original interpretation of quantum mechanics \cite{SchrodingerIV}, in which for a single electron $-e\rho(\vec{x})$ was assumed to be a (classical) charge density. Schr\"odinger wrongly assumed it could be used in Maxwell's equations. There were two reasons for this failure. In the many-electron situation, the density $\rho_j(t, \vec{x}_1, \vec{x}_2,\ldots)$ cannot be used in connection with classical electrodynamics and the not yet invented QED should have been used instead of classical electrodynamics. The take away from Schr\"odinger's attempt is that in 1926 it was appropriate to consider a fundamental distributed quantity. The fundamental significance of $\rho_j(t, \vec{x}_1, \vec{x}_2,\ldots)$ is that it gives where the system is in configuration space, as laid out in EQM 1. Any other interpretation or significance of $\rho_j(t, \vec{x}_1, \vec{x}_2,\ldots)$ should be derived from EQM 1 together with the following postulate and other physical circumstances that can be assumed.

\begin{peqm}
 \label{post:TD} The equation of motion: There is a linear and unitary time development of the state, e.g.,
\begin{equation}
i\hbar \partial_t \Psi = H \Psi,
\label{schroedinger}
\end{equation}
where $H$ is the hermitian Hamiltonian. The term unitary signifies that the value of the left hand side in (\ref{norm}) is a constant of motion for any state (\ref{Psi}) of the system.
\end{peqm}

When investigating how the theory describes the world we observe, the Hamiltonian has to be assumed to have realistic features. We should realize that we have no proper understanding of a world where the interactions are different from those that govern this world. In particular, measurements are physical processes governed by the existing forces. 
As the standard model of particle physics is formulated in terms of locally interacting fields, it will be assumed that interactions are local and that there are locally conserved particle currents.

The following relations lends support to the interpretation of the density (\ref{density}) as the distributed position. For the sake of simplicity, the spin index $j$ and the time dependence are omitted here. 

From the system density (\ref{density}) a single-particle density for the $N$ particles of the same kind can be calculated
 \begin{equation}
  \rho(\vec{x}) = N\int d^3x_2 d^3 x_3 \cdots \rho(\vec{x}, \vec{x}_2, \vec{x}_3, \ldots )
 \label{eq:single_particle_density}
 \end{equation}
 Similarly, a two-particle density can be defined as
  \begin{equation}
  \rho(\vec{x}_a, \vec{x}_b) = N_aN_b\int d^3x_3 d^3 x_4 \cdots \rho(\vec{x}_a, \vec{x}_b, \vec{x}_3, \vec{x}_4 \ldots )
 \label{eq:two_particle_density}
 \end{equation}
 where $N_b = N_a -1$ if the two particles are `identical'.
If relevant collective coordinates are introduced $X_1, X_2, \ldots$, a corresponding density $\rho(X_1, X_2, \ldots)$ can be defined. These densities are physically significant. If all interactions are local, the single-particle density is locally conserved,
 \begin{equation}
  \partial_t \rho + \nabla \cdot \vec{j} = 0,
 \end{equation}
 and so is the two-particle and the system density. 
 
 The following illustrates the physical significance of the single-particle density. 
 For at bound system, the single-particle density can be probed with an external potential by measuring the related energy change,
 \begin{equation}
  \Delta E = \int d^3x \, V(\vec{x}) \rho(\vec{x}) .
 \label{eq:deltaE}
 \end{equation}
 In nuclear and particle physics, electron scattering can be used to extract the ground state charge density, which corresponds to the single-particle density of protons and a combination of the up and down quark single-particle densities, respectively. 
 
 Two-particle and many-particle correlations in the wavefunction of a complex system can be extracted from static properties and excitation probabilities to excited states with various properties. Hund's rule for the structure of the ground-state properties of atoms states that if the valence $(n,l)$ shell is half-filled, then the system has maximal total spin $S$ and orbital angular momentum $L$. In this state, the electrons are as far away as possible from each other, minimizing the electron-electron repulsive Coulomb energy. The corresponding two-particle density is zero for $\vec{x}_a = \vec{x}_b$. 
 
 The structure of molecules is interesting because it illuminates the relevance of both single-particle densities and correlations. The electrons move in the electric field from the nuclei, and, in the Oppenheimer-Born approximation, the nuclei move in the electric field of the electrons given by their single-particle density. The electronic energies give rise to correlations of the positions of the nuclei that are well represented by the structure of the system density (\ref{density}).
 
 Collective coordinates are suited for the location and orientation of macroscopic bodies. The density in those coordinates may describe where the different items in the laboratory are located, which enables the quantum description of experiments. Note that it is the interactions that cause atoms to exist, make them bind together into molecules, crystals, and different kinds of macroscopic bodies. That the macroscopic objects are found in well-defined positions is due to decoherence \cite{joos2013decoherence}. 
 
 Of course, EQM 1 does not affect any dynamical process or which wavefunctions that are allowed. Specifically, EQM 1 does not dictate the answer to the preferred basis. Instead, the answer is given by the decoherence in systems of macroscopic objects surrounded by gases of molecules, photons, neutrinos, and gravitons. EQM 1 is chosen to enable the investigation of how and if, EQM can describe nature. It fulfills an epistemic need.

The discussion above shows that {\em if we wish to interpret the meaning of $\rho_j(\vec{x}_1, \vec{x}_2, \ldots)$ without any attention to the measurement process, the interpretation that is given by EQM 1 or something to the same effect seems unavoidable. }  

\subsection{Alternative postulates}

Below are listed the standard measurement postulates, which are replaced by EQM 1 and 2.
\begin{description}
\item[S 1] The state of a physical system is a normalized vector $| \Psi \rangle$ in a Hilbert space $H$, which evolves unitarily with time.
\item[S 2] Every measurable quantity is described by a Hermitian operator (observable) $B$, acting in $H$.
\item[S 3] The only possible result of measuring a physical quantity is one of the eigenvalues of the corresponding observable $B$.
\item[S 4] The probability for obtaining eigenvalue $b$ in a measurement of $B$ is $P(b) = \langle \Psi | \pi_b |\Psi \rangle$, where $\pi_b$ is the projector onto the eigen-subspace of $B$ having eigenvalue $b$. 
\item[S 5] The post-measurement state is (the result of the unitary development during the measurement of) $\pi_b|\Psi \rangle / P(b)^{1/2}$.
\end{description}

Some modern formulations of the postulates allow for positive operator value measurements, but that generalization offers nothing extra here. It is the same as the projection value measurement postulates S 2-5 up to a unitary transformation \cite{nielsen2010quantum}. 

The standard postulates amount to a partial interpretation of quantum mechanics. It is complete enough for the investigation of well-defined localized systems, but not for the environment as a whole. EQM 1 and 2 imply the content of these postulates. In the comments to EQM 1, it is shown the state belongs to a Hilbert space so that EQM 1 and 2 imply S 1. In EQM, the measurement is as any other process described by the dynamics given by EQM 2. Section \ref{singleMes} shows how S 2 and S 3 are implied.

Everett \cite{everett1957relative}, Wallace \cite{wallace2012emergent}, and others, posit that the wavefunction belongs to a Hilbert space. As they do not explicitly state which Hilbert space they refer to, one might wrongly conclude that it is an abstract Hilbert space with no relation to the physical world. Nevertheless, the dynamical equation (\ref{schroedinger}) relates each degree of freedom with a particle type. Nevertheless, these authors do not give any rule for what the wavefunction amplitude signifies. One cannot directly rely on standard practice, as that is motivated by the standard postulates, in particular, the Born rule (S 4). Wallace argued in connection with the decoherence theory that the Hilbert norm measures the relative importance of different parts of the state. To start with, postulating the mathematics and derive the physics from this, becomes at best, a backward way to define the theory\footnote{The postulation of Hilbert space in S 1 seems equally backward. A normalization requirement is implied by the Born rule, which in turn implies that the state belongs to a Hilbert space. This possibility puts the derivation of the Born rule by Gleason's theorem \cite{GleasonMeasureOnHS} in a new light.}.

Geroch \cite{Geroch1984} suggested a related interpretation of the quantum state, which states that a region of configuration space is `precluded' if the wavefunction is very small there. This suggestion corresponds to ignoring contributions from configuration space where the system is hardly present. From EQM 1, this recipe may be motivated in some situations, but, as with the Hilbert space postulates, the Geroch approach does not give a physical meaning to the wavefunction.

A postulate formulated in terms of momenta rather than the position in configuration space, can that replace EQM 1? That seems not to be a practical starting point for our description of the world. We observe a `classical world' of macroscopic objects at reasonably well-defined positions. The position basis is useful when formulating the decoherence theory, which explains the appearance of the classical world. In section \ref{singleMes}, the meaning of the amplitudes (absolute square) in another basis is derived from EQM 1 by considering experiments in which a physical process unitarily transforms that basis to localized wave packets. To mimic this procedure in the momentum basis would entail processes of which we have no elementary understanding, which is necessary for the argumentation that starts directly from the postulates. To put it simply, if we cannot give a meaning to localized states, how can we then make any connection with the world we observe.

\section{Basics of Measurements \label{singleMes}}

It is difficult to analyze which quantities can be measured based on the general and abstract view of quantum mechanics, which is meant to be valid for any type of interaction. As mentioned above, it is assumed that the fundamental interactions are not only local but that they can give rise to the kind of objects that we have around us, for example, detectors and laboratories.

Detectors can typically record that a particle entered it, which can be used to create position information. The momentum of a charged particle can be transformed into a measurement of position. 
The measurement of photon energy can be transformed into the measurement of a position using a grating. 
The measurement of the angular momentum of an atom can be transformed into a photon energy measurement by the Zeeman effect or position by a Stern-Gerlach apparatus. 
These are examples of measurements of physical quantities which correspond to Hermitian operators and can be transformed to a position measurement. 
Even the recording of the time for an event is, in principle, transformable to a position.

There are measurement methods that do not rely on position measurements. For example, gamma photon energies and other high energy particle energies can be measured by recording the number of produced secondary particles. The primary way in which such detectors are usually calibrated by comparing with position transforming methods. The following discussion of measurements will be confined to the recording of a particle entering a detector, which we can call particle recorders. 
These detectors may be a part of an array of detectors in order to get position information from which detector was hit.

Particle detectors react when a particle is entering a particular volume or area. 
There is an infinite set of states with support inside the volume (area) and another infinite set of orthonormal states with support only outside. Together they make up a complete basis. 
The Hermitian operator that corresponds to measurements with this detector can be defined such that all the inside states are eigenstates with a common eigenvalue and the outside with another value.
This detector can only tell whether a particle came into it or not. 
For a particle recorder array, the Hermitian operator may be constructed by associating the same value for all states inside one particle recorder, but different values for the different recorders. 
Additionally, another value should be attributed to the outside of all particle recorders. 
In summary, this detector records if any and which of the individual particle recorders fired. This record corresponds to a particular eigenvalue of a Hermitian operator.

The detector described so far is highly idealized. For example, it is unrealistic that a particle recording detector can register particles at any energy. 
However, at a specific experiment, the energy range of the particles is limited. 
The described model is relevant as long the efficiency is close to 100\% in the real experiment. 
 
\begin{figure}[htbp]
\begin{center}
\includegraphics{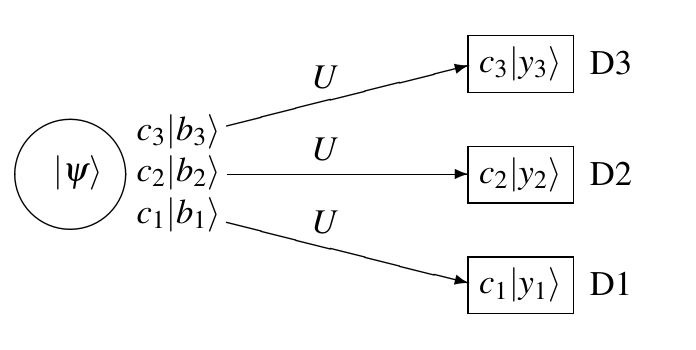}
\caption{The position detector consisting is the particle recorders D1-D3 receive the different components of the wavefunction $|\psi \rangle$ due to the unitary transformation $U$. The state $| a_n \rangle$ transforms to $| y_n \rangle$ by $U$.}  
\label{experiment}
\end{center}
\end{figure}
In figure \ref{experiment}, is shown schematically what is involved when a property corresponding to an operator $B$ is measured with an array of particle recorders. There is physical process that transforms the corresponding eigenstates $|b \rangle$ to become a state corresponding to that one of eigenstates of the detector system $|y \rangle$. If the unitary operator representing the transformation is denoted $U$, we have
\begin{equation}
|y\rangle = U |b\rangle.
\label{astates}
\end{equation}
If the state to be measured is written in the eigenstates to $B$,
\begin{equation}
 |\psi\rangle = \sum_b c_b | b \rangle
\label{psi}
\end{equation}
then the state that enters the position detector system is
\begin{equation}
\sum_b c_b U|b\rangle = \sum_b c_b |y_b\rangle.
\label{detectordist}
\end{equation}
This expresses that the different eigenstates $|b\rangle$ enters separate particle recorders and is there represented by $|y_b\rangle$.

As the functions $y_b(j,\vec{x})$ with differing value of $b$ have disjoint spatial support, the density of the state (\ref{detectordist}) is
\begin{equation}
\rho_j(\vec{x}) = \sum_b |c_b|^2 |y_b(j,\vec{x})|^2 .
\end{equation}
 
It describes where the system is according to EQM 1. Summation over the spin and integration over the volume of one of the particle recorders will give the value $|c_b|^2$, where $b$ is the eigenvalue of $B$ associated with that recorder. 
The interpretation of this result is that 
\begin{equation}
\rho_b = |c_b|^2
\label{rhob}
\end{equation}
as a function of the discrete variable, $b$ tells where the system is for the eigenvalue of $B$.  

The interaction between the interior of a particle recorder and the corresponding state of the particle $|y \rangle$ and the signaling to the environment give rise to decoherence such that branches appear, in which precisely one detector has fired.
There is further discussion of decoherence in section \ref{sec:decoherence}.

In order to simplify the notation, it will be assumed that the state $|\psi\rangle$ (\ref{psi}), instead of the transported state (\ref{detectordist}), directly interacts with the (composite) detector. 
Then, the interaction with the detector $M$ is described by
\begin{equation}
  \big(\sum_b c_b | b \rangle \big) |M_\emptyset \rangle \rightarrow \sum_b c_b | b \rangle' |M_b \rangle.
\label{eq:measurement}
\end{equation}
The detector changes its state from its nothing registered state $ | M_\emptyset \rangle $ to a state $ | M_b \rangle $, which corresponds to that the state $ | b \rangle $ is registered. 
The state of the system before and after the measurement $ | b \rangle $ and $ | b \rangle' $, respectively, can be the same state. 
In reality, there is a set of states of the detector that all correspond to the value of $b$. This and similar complications are henceforth ignored.

According to Everett, the observation process is described by
\begin{equation}
\big(\sum_b c_b | b \rangle' |M_b \rangle \big) |O_\emptyset \rangle \rightarrow \sum_b c_b | b \rangle' |M_b \rangle |O_b \rangle,
\label{eq:observation}
\end{equation}
where the state of the state of the observer $O$ is altered to having observed the value that the detector has measured. 
The distribution $\rho_b$ gives the position of the total system over the branches. Another way to express this, the value of $\rho_b$ gives the presence at the branch with the outcome $b$ of the observer and everything else entangled with the measurement result.

The significance of $\rho_b$ is the most important result from this section, as it is crucial for the derivation of the Born rule. The postulates S2 and S3 have been derived, which is apparent from equations (\ref{eq:measurement}) and (\ref{eq:observation}), which expresses that the detector and its observer measures an eigenstate of the intended operator. This finding is familiar from Everett's original discussions \cite{everett1957relative}, but there the standard Hilbert space rules were used, while the current discussions show how S2 and S3 emanate from EQM1, EQM2, and the assumed properties of the interactions. The locality of interactions and the appearance of locally conserved currents are necessary assumptions for the setup in figure \ref{experiment} to be viable.

\subsection{Decoherence: selector and protector\label{sec:decoherence}}

There may be an ambiguity in the transformation (\ref{eq:measurement}). If $|\psi \rangle$ is written in another basis $|x \rangle$ that are eigenstates to operator $X$, $[X,B] \neq 0$, then we get
\begin{equation}
  \big(\sum_x d_x | x \rangle \big) |M_\emptyset \rangle \rightarrow \sum_x d_x | x \rangle' |M'_x \rangle,
  \label{eq:X}
\end{equation}
where the detector states $|M'_x \rangle$ are linear combinations of the states $| M_b \rangle$.
From (\ref{eq:X}), it might look like as if the quantity $X$ has been measured.
However, the assumed experimental setup, figure \ref{experiment}, with realistic properties of the particle recorders guarantees that $B$ is measured, as expression (\ref{eq:measurement}) suggests. 

When a particle recorder is excited because the system enters it, very many degrees of freedom get excited. The possibility of interference between the terms in the right-hand side of (\ref{eq:measurement}) is then negligible if the measurement setup is appropriately performed. For example, for an interference to be possible between the term that corresponds to particle recorder number 1 in figure \ref{experiment} and particle recorder 2, all the particles of recorder 1 has to be in the same state, and the same has to apply to recorder 2. Considering the vast number of particles that change their state during an interaction with the incoming particle, the presence of this situation is minimal. 

The storing of the measurement data into some memory, constructed to be resilient and with considerable redundancy, is itself enough to hinder any coherence between the possible measurement values. If the data that is written on paper, the ink molecules that attach to the paper are not likely to lose their position by quantum spreading. They attach to the paper and each other, forming macroscopic structures. It is well known that macroscopic structures are measured continuously by their surroundings \cite{joos2013decoherence}. The quantum Zeno effect then implies that the quantum uncertainty of the position of the writing will be minimal. If nothing else protects from coherence between the terms of (\ref{eq:measurement}), the way we store the data guarantees that we will not notice any coherent effects destroying our knowledge about them.\footnote{Vaidman \cite{vaidman2016all} has defined `worlds' as having different macroscopic structure. This definition might not be generally appropriate. However, successful measurement results give rise to different macroscopic structures, so the branches that are discussed here are precisely Vaidman `worlds'.}.

The decoherence defines a unique basis for the detector states. In the case of a measurement setup like in figure \ref{experiment} the states $|M_b\rangle$are local in space, while the alternative basis states $|M'_x \rangle$ used in ($\ref{eq:X}$) are not. The non-diagonal matrix elements of a local operator $L$ within the $|M_b \rangle$-basis are FAPP zero, while in the alternative basis, that is not at all the case. The arguments in the last paragraph imply that for any experimental setup, the branches are kept apart in configuration space, so that a local operator still have FAPP zero matrix elements between different branches%
\footnote{Rovelli wrongly claimed \cite{rovelli1996relational} that ``decoherence depends on which observation $P$ will make'', where $P$ is a second observer. In a laboratory setting, it is the detector system that create decoherence and defines which quantity is measured.}. There is no ambiguity in the measurement basis. 

The derivation of the decoherence mechanism \cite{zeh1970interpretation,joos1985emergence} is based on the traditional interpretation of quantum mechanics. Joos \cite{joos2000elements} and Baker \cite{baker2007measurement} questioned the use of decoherence theory to infer the Born rule as it is already assumed. However, decoherence theory primarily relies on the Born rule to conclude that the environmental particles are `measured' somewhere after they scattered off `macroscopic' systems. The interpretation given by EQM 1 serves well to replace the Born rule in decoherence theory.

Kent \cite{kent2010one} and Dawid and Th\'ebault \cite{dawid2015many} argued that the `fuzziness' that decoherence gives to the branches definition is unacceptable.
The point is that the observer's beliefs can then not give a well-defined theory of a well-defined Born rule. 
However, the fuzziness of branches is the fuzziness of actual measurements. 
It is generally assumed that the experiments can be made arbitrarily exact, in principle. 
If that is not the case, several interpretations will be in trouble. If the fuzziness of decoherence is a problem for EQM, the experimental verification of the Born rule is equally in trouble. The decoherence is an integral part of our understanding of the measurement apparatus and is independent of any interpretation of quantum mechanics. 

If decoherence has not occurred, no measurement has been performed. Consider a photon experiment in which there is no decoherence causing detector. Then the photon will fly around in a maze of mirrors, beamsplitters, etc. Finally, the photon will be absorbed in one of those elements of the setup or a laboratory wall or similar body that cause decoherence without any recording. As long we have a quantum phenomenon with only very few degrees of freedom of elementary or collective nature, it is not yet a measured system.

In the continuation, the term `world' includes all the relevant environment. If we are in a well-defined branch and we have do not consider recoherence, then that is our world.
Branch denotes one of several structures for which the world wavefunction has a negligible amplitude at regions separating them during an extended period.

\subsection{The measurement result}

So far, it has been established that the measurement setup, as in figure \ref{experiment}, can create one well-defined branch for every possible measurement value, which are eigenvalues to a Hermitian operator. The quantum state of the branch is that of the eigenstate (with amplitude $c_b$) entering the detector. The standard postulates S 1, S 2, and S 3, section \ref{postulates}, are fulfilled in each branch, as well as S 5 if the current branch is renormalized to norm one by the observer within a branch\footnote{Zurek \cite{zurek2013wave} proved under the assumption of non-disturbing measurement that the measured basis states have to be orthogonal in order for the measurement apparatus to differentiate them.}. 

Looking at the many branches from the outside, the question ``What reading did the observer get?'' is equivalent to ``What is the distribution of observer readings?'' - 
The answer is given by the distribution $\rho_b$ (\ref{rhob}). 
This value can also be arrived at calculating the total density (the norm) of the $b$-term in the final state of (\ref{eq:measurement}) or (\ref{eq:observation}). Note that once decoherence has taken place, the created branches evolve independently, keeping their norms conserved. 

If $\rho(\vec{x}_1,\vec{x}_2, \ldots)$, describes what exists, is ontic, then we must consider that $\rho_b$ is also ontic. This relation implies that in any basis, the distribution $\rho_b$ gives information about what exists, but in general, it is not the full information.
 
\section{Repeated Measurements\label{section:repeated}}

Suppose the detector is able to record several subsequent measurements of identically prepared systems (\ref{psi}). Further, assume that the way the detector interacts with the next system is not essentially affected by previous measurements. The second measurement is described by the transition 
\begin{eqnarray}
\big(\sum_{b_2} c_{b_2} | b_2 \rangle \big) \sum_{b_1} c_{b_1} | b_1 \rangle' |M_{b_1} \rangle \rightarrow \nonumber \\ 
\sum_{b_1b_2} c_{b_2} c_{b_1} | b_2 \rangle' | b_1 \rangle' |M_{b_1b_2} \rangle.
\end{eqnarray}
When the interaction with the observer is included the final state becomes
\begin{equation}
\sum_{b_1b_2} c_{b_2} c_{b_1} | b_2 \rangle' | b_1 \rangle' |M_{b_1b_2} \rangle |O_{b_1b_2} \rangle.
\label{eq:commonState}
\end{equation}
Each sequence of readings belong to different branches. The distribution of observer reading sequences is now\begin{equation}
\rho_{b_1b_2} = |c_{b_1}|^2|c_{b_2}|^2.
\end{equation}
After $N$ measurements, the sequences of observer readings are distributed according to
\begin{equation}
\rho_{b_1b_2...b_N} = |c_{b_1}|^2|c_{b_2}|^2\cdots |c_{b_N}|^2.
\label{eq:rho_rep}
\end{equation}
To focus on the value $b = u$, denote the summed density of all the other values of $b$ by 
\begin{equation}
\rho_{\neg u} = \sum_{b\neq u} |c_b|^2
\end{equation}
and $\rho_u = |c_u|^2$. The sum of the densities (\ref{eq:rho_rep}) over all sequences where $b=u$ appears precisely $m$ times out of $N$ measurements is
\begin{equation}
\rho(m \!:\! N | u) = \frac{N!}{(N-m)!m!}(\rho_u)^m (\rho_{\neg u})^{N-m}.
\label{eq:a_ m-times}
\end{equation}
This gives the total summed density of the branches in which the value $u$ was found by the observer $m$ times. 
Hence, the question ``How many times have the observer measured the value $u$?'' is answered by $\rho(m\!:\! N | u) $ as a distribution over $m$-values.  

For large number of measured systems $N$, the distribution (\ref{eq:a_ m-times}) may be approximated by a gaussian, see Feller \cite{feller1968introduction},
\begin{equation}
\rho(m\!:\! N | u) \approx \frac{1}{(2\pi N\rho_u\rho_{\neg u})^{1/2}} \exp\big(-\frac{(m-N\rho_u)^2}{2N\rho_u\rho_{\neg u}}\big).
\label{mGaussian}
\end{equation}
The distribution (\ref{mGaussian}) may be represented as function of the relative frequency $z=m/N$ taken as a continuous variable. 
The properly normalized position or presence distribution with respect to $z$ is
\begin{equation}
\rho(z | u) = \big( \frac{ N }{ 2\pi \rho_u\rho_{\neg u} } \big)^{1/2} \exp\big(-\frac{ N(z-\rho_u)^2 }{2 \rho_u\rho_{\neg u} }\big).
\label{eq:relative}
\end{equation}
As $N \rightarrow \infty$ this density approaches the delta function $\delta(z-\rho_u)$. 
This relation says that at infinitely large $N\,$, there is only one value of the frequency $z=\rho_u$. 
It might look like a big stride towards proving Born's probability rule, but $\rho(z | u)$ is an approximate result. 

To get from the exact expression for $\rho(m: N\, |\,u)$ (\ref{eq:a_ m-times}) to the continuous frequency distribution, the interval $[0,1]$ is divided into a set of intervals $\{I_k\}$,
\begin{equation}
I_k = [0,1] \,\cap [z_k-\Delta z/2, \, z_k+\Delta z/2[ , \, z_k = \rho_u + k\Delta z.
\end{equation}
The index $k$ belongs to the minimal set of integers such that $\{I_k\}$ covers $[0,1]$. 
Define $\tilde{\rho}(k)$ as the sum of densities $\rho(m: N\, |\,u)$ with $m/N$ in the interval $I_k$. 
Set 
\begin{equation}
\rho_{\Delta z}(z|u) = \tilde{\rho}(k)/\Delta z\; \mbox{if}\; z \in I_k.
\end{equation}
This is a histogram type piece-wise constant function. 
If $\Delta z = \Delta z_1/N^{-1/2}$ and $\Delta z_1$ is small and $N$ is large, then $\rho_{\Delta z}(z|u)$ can be arbitrarily close to $\rho(z | u)$. 

In order to adequately justify the use of the frequency distribution (\ref{eq:relative}), an operator should be found that is closely related to this distribution.
The first guess may be the frequency operator 
\begin{equation}
F_N = \frac{1}{N} \sum_{i=1}^N f_i 
\label{eq:frequency_operator}
\end{equation}
where $f_i$ operates on the $i$-th system being measured with $f|u\rangle=|u\rangle$ and $f|b\rangle = 0$ if $b\neq u$. 
The eigenvalues of $F_N$ are $z = m/N, \, m= 1,...,N$. 
The density related to $F_N$ acting on this state is given by (\ref{eq:a_ m-times}) with $m$ replaced by $zN$. 
As pointed out by Squires \cite{SquiresOnProof}, the density values of this discrete distribution 
approaches zero as $N \rightarrow \infty$. 

Take instead, the operator $F_{N\Delta z}$ defined by its action on products of eigenstates to the operator $B$. If the frequency of the eigenvalue $u$ is in the interval $I_k$ with midpoint $z_k$, then
\begin{equation}
F_{N\Delta z} |b_N\rangle|b_{N-1}\rangle ... |b_1\rangle = z_{k} |b_N\rangle|b_{N-1}\rangle ... |b_1\rangle.
\end{equation}
The density of this operator is $\tilde{\rho}(k)$. As the eigenvalues $z_k$ of $F_{N\Delta z}$ is a discrete set its density distribution $\rho_{z_k} = \tilde{\rho}(k)$ is represented be a bar graph rather than the histogram that represents $\rho_{\Delta z}(z|u)$. 

To see the behavior of these densities as $N$ approaches infinity, the Chebyshev inequality \cite{feller1968introduction} can be applied to the distribution $\rho(m\!:\! N\, |\,u)$ (\ref{eq:a_ m-times}). 
The result can be written as  
\begin{equation}
\sum_{|m/N-\rho_u| > \Delta z/2} \rho(m\!:\! N\, |\,u) \leq \frac{4\rho_u\rho_{\neg u}}{(\Delta z)^2 N}.
\label{eq:chebyshev}
\end{equation}
From this follows that $\sum_{k \neq 0}\tilde{\rho}(k) \rightarrow 0$ as $N \rightarrow \infty$. Hence, $\tilde{\rho}(0)$ approaches one for any fixed value of $\Delta z$. 
The delta function limit of $\rho(z | u)$ is confirmed by the exact calculation.

The quantity $\rho(z|u)$ is a continuous approximate representation of $\rho(m\!:\! N | u)$, which is a sum of the densities of several branches. 
The interpretation is that $\rho(z|u)$ gives the position distribution for the relative frequency $z$ of everything entangled with the measurement result. The presence of the observer within an interval in the relative frequency of length $dz$ is $\rho(z|u)\,dz$.   

\section{The Born Rule \label{bornsrule}} 

The Born rule is an indispensable tool when investigating the quantitive features of microscopic systems. It relies on the concept of probability, but this concept is not straightforwardly available in EQM. All branches with non-zero amplitudes are created, so there is no single outcome about which we can be uncertain. 
Classical probability theory is silent about the relative frequency of a particular outcome that the observer should expect to see. Nevertheless, an observer will in a particular branch have seen a more or less random sequence of outcomes, and the observer's presence distribution (\ref{eq:relative}) seems to say that the Born rule is valid. Though the notion of classical probability is not valid here, there may be another concept available. Denote that hypothetical quantity Everettian quantum probability, or only quantum probability when EQM is assumed.

Papineau \cite{papineau1996many} analyzed what requirements quantum probabilities need to fulfill. He argued that it is sufficient for quantum probabilities to relate to `non-probabilistic facts’ in the same way as probability does in practice. He identified the two ways in which this relation exists. First, he identified the `Inferential Link,’ which is the use of observed frequency from a finite number of repetitions for the inference of a value of a probability. Second, Papineau identified that we use probabilities to guide decisions, which he called the `Decision-Theoretic Link.’ If the two links can be established in EQM, Papineau's analysis makes away with the idea that the notion of probability necessitates the existence of uncertainty. No doubt, uncertainty is indispensable in the context of single outcome probabilities, for which the term `classical probability' is used here. For Everettian quantum probabilities, uncertainty is neither available nor necessary.

In section \ref{sec:statistics}, starting from the result of section \ref{section:repeated} it is shown that the inferential link is present. Inferring the value of $\rho(u)$ from observations works equally well in EQM as in a single outcome interpretation. Section \ref{brancher} is devoted to showing that the work by Greaves and Myrvold \cite{greaves2010everett} combined with the results of sections \ref{section:repeated} and \ref{sec:statistics} give the decision-theoretic link. 
 
\subsection{Frequentist and Bayesian probabilities\label{sec:proving}}

The proof of the inferential link is closely related to {\em the frequentist view of probabilities, where the probability is the relative frequency from infinitely many repetitions} \cite{feller1968introduction,feller2008introduction}. 
The following criticism against the frequentist view of classical probability, adapted from Appleby \cite{appleby2005facts,appleby2005probabilities} and Wallace \cite{wallace2012emergent}, should be responded before accepting a proof relying on frequencies observed after many or infinitely repeated events. 
\begin{enumerate}
\item \label{no-probability-def} C: After infinitely many repetitions, the value of the relative frequency can deviate from the probability. 
The probability of such sequences tends to zero, but this latter use of the notion of probability makes this definition of probability circular. 

R: In EQM, all possible sequences of measurement results together constitute the reality. Not only a single sequence as in the case of a single outcome at each measurement. After infinitely many repetitions, the universal wavefunction is only located at the relative frequency $z=\rho_u$. The observer sees a random sequence that suggests an analysis in terms of probability, which, given the behavior of the presence distribution, will be taken to be $\rho_u$. This analysis creates no circularity as presence is a quantity on its own, not derived from the probability concept. However, the probability concept does not directly appear from such an analysis. As pointed out by Caves and Schack \cite{caves2005properties}, this deficiency was the problem with the attempts by Saunders and others that have tried to use Finkelstein's finding that 
\begin{equation}
 \| (F_N - |c_u|^2) \prod_{k=1}^N |\psi \rangle_k \| \rightarrow 0 \; \mbox{as} \; N \rightarrow \infty,
\end{equation}
to prove the Born rule. If the Hilbert norm does not have any physical meaning, there is no meaning to the members of the sequence off in which the limit is taken. The criticism against the otherwise mathematically correct analysis of Hartle \cite{hartle1968quantum} and Gutmann \cite{gutmann1995using} was essentially the same. These authors showed that the state of infinitely many identical systems $ |\Psi_{\infty} \rangle = \prod_{k = 1}^{\infty} | \psi \rangle _{k}$ is an eigenstate to the frequency operator $F_{\infty} = \lim_{N \rightarrow \infty}F_N$. These proofs had to battle with the difficulties of non-separable Hilbert spaces. The lack of a proper interpretation of the wavefunction, such as EQM1, causes those derivations to be mere mathematical exercises. Nevertheless, they do add to the consistency of the $N \rightarrow \infty$ limit of equation (\ref{eq:chebyshev}).

\item \label{infinite-sequence} C: It is impossible to make infinitely many repetitions. 	

R: In a theoretical analysis, it is possible to consider thought experiments in which there are infinitely many repetitions.
\item C: There is no well-defined frequency of a particular outcome in an infinite sequence, as a reordering can change the value.

R: The universe of all of the branches is left unchanged under reordering if all branches are reordered in the same way with respect to their ordinal number in the sequence. Any other reordering would violate the branches being the result of repeated experiments.
\item C: Any particular infinite sequence has zero probability, so how can those with the `right frequency' be favored against the one with another frequency?

R: This is reminiscent of the observation by Squires \cite{SquiresOnProof}, that for any value of $m$ the density $\rho(m\!:\! N | u)$ (\ref{eq:a_ m-times},\ref{mGaussian}) approaches zero as $N$ approaches infinity. As was seen above, by bunching together all sequences into intervals in relative frequencies, the probability of the 'right frequency' interval approaches one.

\item C: The first (finite) part of an infinite sequence is a vanishingly small compared with the rest and give no reliable information about the infinite sequence. 

R: This is the problem of making statements with certainty from an observed sequence. This problem implies that we, at best, can learn about the sought frequency with some probability. Thus, there has to be something more to the notion of probabilities than frequencies. Probabilities must also relate to the beliefs of agents that are uncertain about actual realities. The frequentist must then answer why it is not only a matter of beliefs, as is the case in the subjective Bayesian view of the probability concept. In the case of a deterministic process, an agent's probability assertion seems to merely reflect the agent's subjective knowledge state and her corresponding assessment of the process. However, Lewis \cite{lewis1980subjectivist} opened for the possibility that even if the subjective view is the primary view of probabilities, there may still be possible that, in some instances, there can exist objective probabilities. If they do not exist in the case of a deterministic processes in which an agent is in principle able to know all the facts, then objective probabilities seems to require that there are hidden variables that are impossible to access fully, a fundamental randomness, or, as in EQM, that it is an illusion that only one alternative happens.  

\end{enumerate}

The frequentists view is that, under the given circumstances, the probabilities are objective properties. In EQM, the wavefunction closely describes a real objectively existing physical system. Correspondingly, finding the Born rule in EQM entails the finding of an objective property.

As mentioned above, an alternative to the frequentist view of probabilities is the subjective Bayesian view\footnote{There is also an objective Bayesian view of probabilities, where objective refers to that we all might agree on the probability value. Here we also classify such probabilities to be subjective.}. 
In this view, probabilities reflect an agent's estimate of the likelihood of a particular outcome. 
De Finetti \cite{de2017theory} and Savage \cite{savage1972foundations} has advocated this understanding of probability. 
The theory concerns the beliefs of rational agents for which Savage identified the requirements formulated as postulates. These postulates, which give a foundation for subjective probability theory, are formulated in terms of decisions by rational agents. Greaves and Myrvold \cite{greaves2010everett} have reformulated these postulates to suit the situation of quantum measurements. This formulation turns out to be useful to establish Papineau's decision-theoretic link.

Most notable in this theory is the update of probabilities a rational agent will make on the discovery of new information,
\begin{equation}
	P(A|B) = \frac{P(A\cap B)}{P(B)} = \frac{P(B|A)P(A)}{P(B)}.
\label{eq:bayesupdate}
\end{equation}
Here, $P(A|B)$ is the probability of $A$ when the agent know $B$ to be true, $P(A\cap B)$ is the agent's probability for both $A$ and $B$, and $P(B)$ is the total probability for $B$. 
This expression originates from the identity $P(A\cap B) = P(A|B)P(B)=P(B|A)P(A)$. 
In order to evaluate the update expression, the agent has to analyze the process that leads to an outcome. The strength of the Bayesian view is that it clarifies the notion of probability.

The theory is only a skeleton, which has abstracted away the world about which the agent has beliefs. 
In order to get any values of the probabilities, the nature of the world has to be taken into account by the agent. 
The beliefs are about some features of the world around us. 

\subsection{Statistics and single-outcome believer\label{sec:statistics}}

Consider an observer who believes there is no branching, only one outcome at every time.
She is recording the results from a well-designed measurement process of a quantum phenomenon where the state contains more than one possible value. 
After a long sequence of measurements, according to EQM, the observer is distributed over very many branches. In each branch, a random sequence is observed, which calls for statistical analysis by the observer. The observer will assume that there is a probability $P_u$ of measuring the value $u$ in a single measurement. The probability of the measured relative frequency $z$ after $N$ repeated measurements, for this value of $P_u$, is then
\begin{equation}
P(z | u) = \big( \frac{ N }{ 2\pi P_u(1-P_u)} \big)^{1/2} \exp\big(-\frac{ N(z-P_u)^2 }{ 2P_u(1-P_u) }\big).
\label{eq:Prelative}
\end{equation}
As this is a very narrow distribution for large $N$, a frequentist analysis would give that $P_u$ is in some narrow interval around the measured value of the relative frequency, with some low $p$-value. The Bayesian analysis, assuming de Finetti's infinite exchangeability, gives rise to the probability distribution for the value of $P_u$ conditioned on the measured relative frequency $z$,
\begin{equation}
  P(P_u | z) = \frac{P(z | u) P(P_u)}{ \int_0^1 \! dP_u \, P(z | u)P(P_u)}.
  \label{eq:bayesianupdate}
\end{equation}
Here, $P(P_u)$ gives how likely the observer believed different values of $P_u$ were before the observation was made. If it is constant, as may be the case if there was no previous information, the dependence of $P(P_u | z)$ on $P_u$ will be given by $P(z|u)$.

\begin{figure}[!htb]
\begin{center}
\includegraphics[scale=0.45]{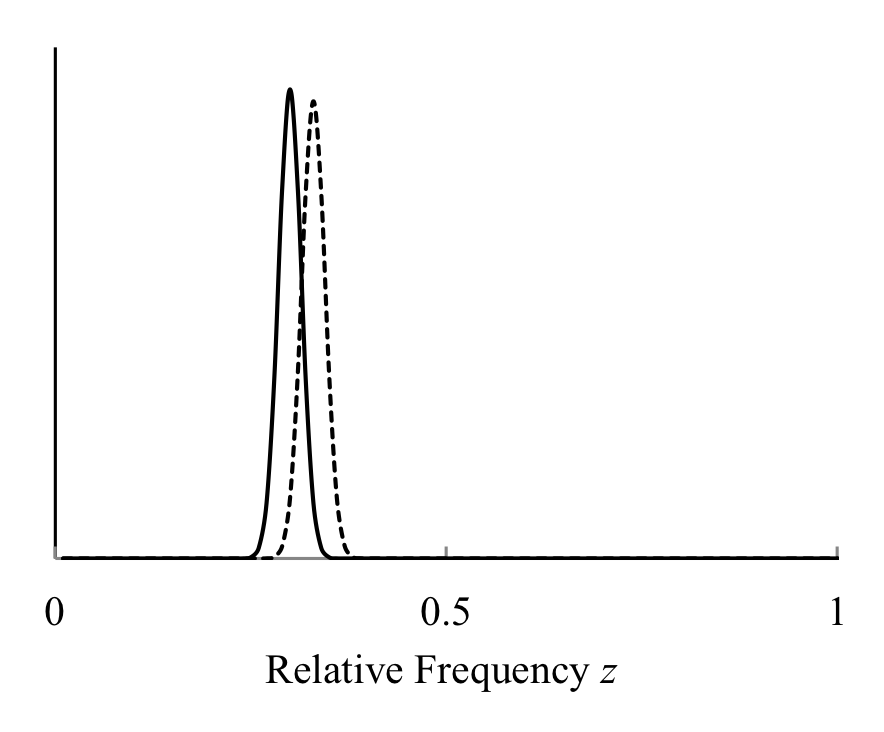}
\caption{\label{fig:binomials}The solid line shows the density, the presence distribution, $\rho(z | u)$ (\ref{eq:relative}) for $\rho_u = 0.3$ and $N=1000$. 
The dotted line shows where an observer in a typical branch may estimate the probability $P(z | u)$ to be from the observed sequence alone.}
\end{center}
\end{figure}

The relative frequency $z$ is distributed over all branches according to (\ref{eq:relative}), see figure \ref{fig:binomials}. Hence, the distribution of $P_u$ over the branches may be seen as the folding of the two distributions (\ref{eq:relative}) and (\ref{eq:Prelative}).

As the number of repeated measurements $N$ grows, the width of the probability distribution $P(z|u)$, tends to zero as does the position distribution $\rho(z|u)$. 
After a large number of repeated measurements, the observer sees a relative frequency close to $\rho_u$, and the value of $N$ implies that the value of $P_u$ is probably close to the observed frequency. Hence, the observer believes that the probability $P_u$ is very close to the value of $\rho_u$. 

To summarize, the observer distribution in relative frequency (\ref{eq:relative}) is narrowing in precisely the same fashion as for a classic probability (\ref{eq:Prelative}). 
The integral of $\rho(z|u)$ is dominated by the peak, which implies that the observer's position is mostly where the relative frequency is close to $\rho_u$. 
If the observer believes in a single outcome interpretation, the observer has arguments for the statistical analysis. This observer's position is dominantly, where she has reason to conclude that the Born rule is correct and subsequently uses it to make inference about the wave function. According to EQM 1, where do we expect to find our selves? If we are to expect anything, our expectation of being near the peak of $\rho(z|u)$ will be high, and our expectation of being in the far tails will be low. 
Our expectation agrees with observation, {\em physicists believe in the Born rule}. 

When the standard interpretation quantum mechanics is verified on bases of the Born rule, the data is compared with the expectation we have from the Born rule. As has been demonstrated, EQM gives rise to the same expectation as the Born rule. This finding implies that EQM is equally well verified as quantum mechanics with the standard postulates.

The discussion in the present subsection reveals that EQM supports an agent's inference of wavefunction properties, at least qualitatively. Papineau's inferential link is essentially established. An observer's rational expectations have been used, but without a well-defined probability theory, the expectations cannot be discussed in quantitative terms. The concept of probabilities is not really at hand yet, but the decision-theoretic link will supply that. 

\subsection{Decision theoretic probabilities\label{brancher}}

When all alternatives with non-zero amplitudes are going to be present, it becomes a challenge to understand the appearance of a probability concept, quantum probability, The observer knows that every possibility with non-zero amplitude is represented in some branch. When observing the outcome, she will also branch, each of her `descendants' seeing the value of that branch. 
As mentioned above, Papineau has argued that it is enough to show the inferential link, which was shown in the previous section \ref{sec:statistics}, and the decision-theoretic link, which will be discussed in this section.
Although there is no uncertainty, thus no classical probability, the situation before the measurement warrants much the same decision theory as when classical uncertainty is at hand. 

Deutsch \cite{deutsch1999quantum} pioneered the use of decision theory to understand Everettian quantum probabilities. Wallace \cite{wallace2002quantum,wallace2007quantum,wallace2010prove,wallace2012emergent} and Greaves \cite{greaves2004understanding,greaves2007probability} have continued this work in their own directions. Both base their analysis on the postulates that Savage formulated \cite{savage1972foundations} to define the theory of classical probability. These axioms imply that the decisions a rational agent makes correspond to maximizing the expected utility, 
\begin{equation}
  \langle U \rangle = \sum_A P(A)U_A.
\end{equation}
Here $P(a)$ is the probability, with $\sum_i P(A) = 1$, and $U_A$ is the numerical value of the utility that the agent will get on outcome $A$. 
From considering the decisions the agent will make under a variety of situations, and a variety of utilities the agent might get, her subjective probabilities $P(A)$ are uniquely determined, and the utilities $U_A$ are determined up to an affine transformation. It is not assumed that the agent consciously optimizes the expected utility, but rational behavior implies it nevertheless.

Lewis \cite{lewis1980subjectivist} acknowledges that even if probabilities are primarily subjective, there are instances like radioactive decay where probabilities are objective features. He formulated the link between subjective probabilities and objective probabilities in the Principal Principle. It implies that an agent who knows that there is an objective probability $P$ sets her subjective probability to $P$. The following analysis as well as that of the previous section follows Lewis view, there is an objective feature, the position distribution, that causes a rational agent to have certain expectations.

Wallace has developed Deutsch attempted a proof of the Born rule into an almost acceptable proof. He has constructed a set of axioms, which he claims any rational agent, who believes in EQM, necessarily obeys. The axioms are not self-evident and not sufficiently motivated in his very general setting, but given the axioms, the Born rule follows. Greaves has taken a skeptical attitude against Wallace's attempts and confined her work towards understanding the concept of probability. 

Greaves and Myrvold \cite{greaves2010everett} have reformulated Savage postulates for rationality, to suit the case of experiments performed in branching as well as non-branching situations. The following quote formulated the purpose of their investigation. -- ``The problem is not one of deriving the correct probabilities within the theory; it is one of either making sense of ascribing probabilities to outcomes of experiments in the Everett interpretation, or of finding a substitute on which the usual statistical analysis of experimental results continues to count as evidence for quantum mechanics.'' 

Following Savage analysis, Greaves and Myrvold arrived at an expression for an agent's expected utility. A rational agent will seek to maximize the expected utility\footnote{The $\langle \, \rangle$ notation can be read as a classical probabilistic expectation value or a quantum (presence) average.},
\begin{equation}
\langle U \rangle = \sum_b w(b) U_b,
\label{eq:averageutility}
\end{equation}
where $U_b$ is the numerical value of the utility the agent gets at the outcome $b$. 
The $w(b)$ is a weight that the agent assigns to the outcome $b$. In the case of a single outcome, it is the agent's subjective probability, or credence, of the outcome.  
In the case of a branching universe, Greaves and Myrvold call it quasi-credence.
Fundamentally, it is a subjective property, precisely as the probability is considered to be.
In both cases, the values of the weights are well-defined if the agent is rational. The weights are subject to the condition $\sum_b w(b) = 1$. 

If new information is presented, they are updated according to the Bayesian update expression, 
\begin{equation}
  w(c|b) = \frac{w(c\wedge b)}{w(b)}.
  \label{eq:update}
\end{equation}
The value of $w(c|b)$ give the agents belief or weight of outcome $c$ in a measurement under the condition that $b$ has been measured, while $w(c\wedge b)$ is the weight for the outcome $b$ and $c$.
One particular updating situation is for the particular value that has been measured. 
 If the concept of classical probability applies, the probability of seeing that value is updated to one. 
 According to EQM, after the branching, the observer's descendants experience as if its branch is the world. After the measurement value is known, the rational agent/observer will set the corresponding quasi-credence to one, $w(b|b) = 1$.
 Likewise, after branching, the agent will normalize the quantum state of her specific branch to have amplitude one, as that is her `system' now.
 
For repeated identical independent events, the de Finetti's infinite exchangeability property can be used, which will give the same Bayesian statistical analysis of the branching world as for the non-branching. For example, the agent will attain a weight distribution $w(w_u|z)$ for the single event weight $w_u$ corresponding to the expression (\ref{eq:bayesianupdate}).  
Greaves and Myrvold argue that de Finetti's theorem, contrary to de Finetti's position, provides us with the notion of objective weights. In the case of a single outcome, they are called chance, while in the branching world, Greaves and Myrvold use the term branching-weights. The $w(w_u|z)$ is the subjective weight distribution for what the objective value of $w_u$ might be, given a measured value of the relative frequency $z$.

Greaves and Myrvold assumed that the EQM Born rule was already proven, which gives the branching-weights equal to $\rho_u$. However, that is not assumed in the present discussion. What is then the significance of the weights, $w(b)$, when the world is branching? As for classical probabilities, the weights are given by the agent’s understanding of the world. An agent that interprets the physical world, according to EQM1, will take into consideration its statement about how the current world (branch) will become distributed over any new branches. For simplicity, assume that the agent is sure about which wavefunction is to be measured. The additional complication related to a mixed initial state is trivial to handle once the pure state situation is understood.

In the single outcome case, the classical probabilities correspond to the agent's beliefs about where she will be present after the measurement. In the branching case, the agent knows from EQM 1 how her presence will be distributed. There is here a similarity between the branching and the single outcome cases. This similarity suggests that in the branching case, the weights $w(b)$ should be equal to the presence values, $\rho(b)$. This identification becomes evident in the case of many repeated measurements.

Consider a measurement that give two branches, $u$ and $\neg u$ with weights $w_u$ and $w_{\neg u}$, respectively. Assume that the utility of a branch only depends on the number of times the value $u$ and $\neg u$ has been measured. The total expected utility after $N$ repeated measurements is then
\begin{eqnarray}
  \langle U \rangle_N = \sum_{m=0}^N w(m \!:\! N |u)U(m,u,N-m,\neg u),
  \label{eq:refw}
\end{eqnarray}
where $U(m,u,N-m,\neg u)$ is the utility of a branch in which $u$ appears $m$ times and $\neg u$ $N-m$ times and 
\begin{equation}
  w(m \!:\! N |u) = \frac{N!}{(N-m)!m!}(w_u)^m (w_{\neg u})^{N-m}.
\end{equation}
The multiplicative form of weights of branches after multiple independent branchings can be seen from the expression (\ref{eq:update}), with $w(c|b) = w(c)$ independent of the value of $b$. 

\begin{figure}[!htb]
\begin{center}
\includegraphics[scale=0.45]{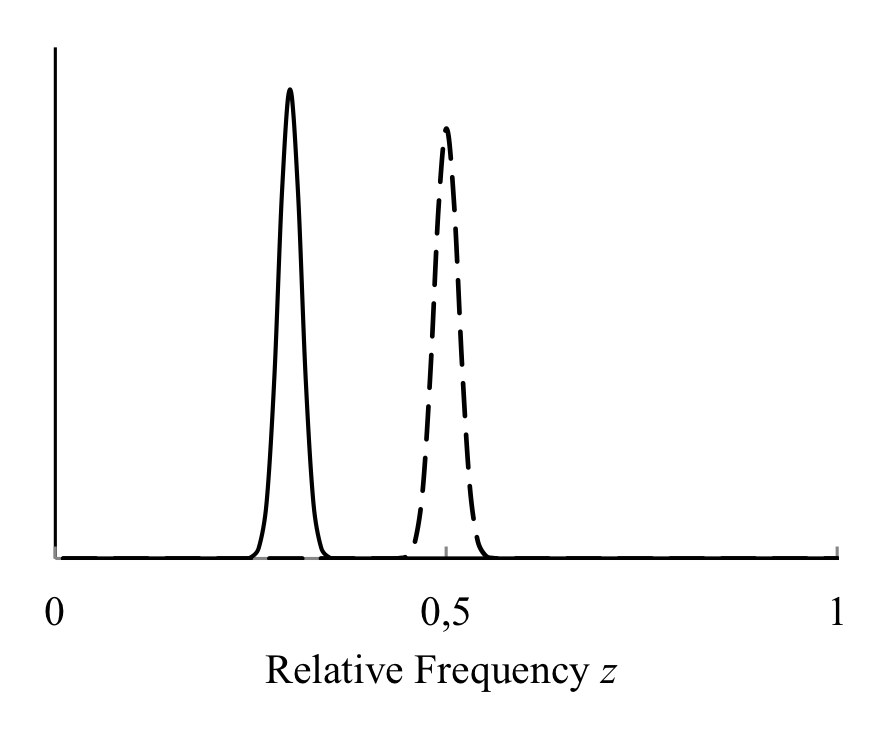}
\caption{\label{fig:wrho}The solid line shows the density, the presence distribution, $\rho(z | u)$ (\ref{eq:relative}) for $\rho_u = 0.3$ and $N=1000$. 
An agent that assumes the weight $w_u = 0.5$ will make decisions such that she gets her favorable utilities where the dashed line is large. With $w_u = 0.5$ she risks getting unfavorable utilities at the position she should expect her self to be according to EQM.}
\end{center}
\end{figure}

From $w(m \!:\! N |u)$ a frequency distribution of the weights $w(z|u)$ is arrived at in the same way as $\rho(z|u)$ (\ref{eq:relative}). 
The functional form of $w(m \!:\! N |u)$ and $w(z|u)$ are identical to that of $\rho(m \!:\! N |u)$ and $\rho(z|u)$, respectively. 
 A rational agent that believes in EQM will have to put $w_u$, the weight of branch $u$, equal to $\rho_u$, the presence of branch $u$, at least for the wavefunction she thinks is the likely one. 
 If she puts her weight $w_u$ different from $\rho_u$, then for large $N$ values she will hardly have any presence where she expects to have most of her presence, see figure \ref{fig:wrho}.  
An agent should make decisions to optimize the utility where she will typically be in the future. That is, the agent should optimize the expected utility (\ref{eq:averageutility}) with $w(b) = \rho_b$.

 What about a single measurement where no branch dominates the presence distribution. The previous argumentation is even then applicable, as the weight the agent should apply to a branch should not depend on whether the measurement is repeated or not. This conclusion follows from the physical picture of local interactions, local detector systems, preparations so that later measurements can be performed sufficiently independent of the previous ones. These are the physical assumptions that were already stated in section \ref{postulates} and which equation (\ref{eq:refw}) rests. 
We have to conclude that a rational agent that knows the wavefunction to be measured and thus knows the $\rho_b$-values is compelled to put her weights equal to those values, $w_b = \rho_b$. Thus the decision version of the Born rule has been proven. 

The reader might feel hesitant at this point, but you might remember your hesitation about classical probabilities before you became educated on that subject. Suppose an uneducated person is offered a choice between a bet which pays twice the punt if event A happens and a bet which only pays 1.5 the punt on B happens. The person is told by an educated friend that there is only a 3 out of 10 chance for A to happen, but 7 out of 10 for B to happen. The uneducated might now reason: B may happen, but it is also possible that A happens, and I get more money in that case. We are not going to do this ten times, and anyhow I am told that even if this thing is repeated ten times, A might happen in all of the ten repetitions. Both A and B are possible outcomes, and no one can deny that, and I will win substantially more in case A. I will rather bet on A. It is only the educated and long term perspective that makes one go for bet B instead. We know that we will find our selves in many decisions, where probabilities apply. If we always stick with carefully estimated probabilities and the corresponding rational decision, we will with high probability be winners. Likewise, an agent confronted with possible choices that will affect the utilities at the different branches can, of course, neglect the weights, if she does not see any relevance in them. However, when the estimate of where she will be in the future after thousands of events is compared with what an equal weight strategy corresponds to, see figure \ref{fig:wrho}, then it becomes clear that the rational behavior is to take into account the weights $w_b=\rho_b$ in the decision as if they are probabilities.

The decision-theoretic link is established. The weights $w_b$ enter into branching world decision making in the same way as classical probabilities do in single outcome interpretations. The Born rule gives the objective values, and they can be estimated using statistics in the same as classical probabilities can be estimated. This result implies that the inferential link is present, not only at the qualitative level as in section \ref{sec:statistics}, but also quantitatively. As the weights behave as probabilities, they deserve to be called probabilities. To distinguish them from their classical counterparts, the term (Everettian) {\em quantum probabilities} is more precise. 
 
\subsection{Explanation and defense of the Greaves-Myrvold theory\label{responses}}

The astonishing and novel feature of the works by Papineau and Greaves is that the operational aspects of probability, statistics, and decision making are available for EQM agents, though they have no uncertainty about the future. Saunders and Wallace \cite{saunders2008branching},\cite{saunders2010chance} have tried to argue for a possible way of `talking,’ a semantics, by which the agent is uncertain about the outcome before a quantum measurement is performed. Such semantics seems contrived and in conflict with the mathematical expressions. The state before the measurement $ | \psi \rangle | M_{\emptyset} O_{\emptyset} \rangle $ contradicts their description in terms of already existing branches. In EQM, all that exists is represented by the wavefunction. Any physical partition has to correspond to an expansion of $ | \psi \rangle$ in some basis $ | a \rangle $. Such an expansion cannot be understood as a partition into any pre-existing worlds, as the individual basis states $ | a \rangle $ are in disagreement with the prepared state, $ |\psi \rangle$. Any description in non-mathematical language that is not supported by the mathematical expressions will produce an ill-defined description of nature. The appearance of probability without any uncertainty being present is exciting progress that should be acknowledged.
 
Kent \cite{kent2010one} criticized Savage's postulates on which Greaves and Myrvold based their considerations.
He argued that many possible strategies are in conflict with the postulates, but are rational. Kent lists seven alternative strategies that do not conform to Savage postulates. 
He claimed that his alternative strategies are rational, but he only argued that those strategies could be applied consistently, which is not the same as rational. It is possible to be irrational and consistent. Two of the strategies are undefined, but the others can easily be shown to be irrational by varying the set of rewards the agent will get.

With the intent to criticize the concept of branch-weight, 
Kent suggests five different computer-generated branching worlds CBU$_{1-4}$ and CBU-qualia. Kent claims that Greaves' and Myrvold's weights do not apply to them. Kent's analysis of these worlds seems insufficient to warrant his conclusion. Anyhow, the present analysis has shown the applicability of the Greaves-Myrvold theory for EQM.

Kent's conclusion from Greaves' and Myrvold's work is that ``Everettians cannot give an explanation that says that all observers in the multiverse 
 will observe confirmation of the Born rule, or that very probably all observers will observe confirmation of the Born rule.'' 
 Indeed, in EQM, there will be some branches with a low presence where the statistics disconfirm the Born rule. 
 However, in a single-outcome interpretation, the Born rule implies that there is a finite probability that we will fail to confirm the Born rule. In kinds of interpretation, we are in the same predicament. Our understanding of the world might be wrong because we have only experienced low probability or low weight events. Independent of the interpretation, we have to assume that this is not the case. EQM gives that the total presence of the branches in which we should have seen the Born rule is overwhelmingly large. Thus we expect to be in a branch where the statistics are in agreement with the Born rule. For a further argumentation against Kent's criticism, see \cite{duhamel2011guildenstern}.
 
Price \cite{price2010decisions} is skeptical towards the existence of `probability' in situations where there is no uncertainty present. As has been shown above, there is no real uncertainty, but there is a distribution situation to which Savage decision theory is applicable. Uncertainty turns out not to be a requirement as the concept presence successfully replaces the classical concept of probability.

Further, Price erroneously regards a person's descendants in the different branches, as if they are different persons. In the example `Legless at Bondi beach,' he discusses the misfortune that swimmer's choice causes to one of his descendants as if the swimmer caused harm to another person in an unethical way. However, the choice corresponds to a gamble that, in a single outcome scenario, could, if unlucky, give a disastrous result. There is no reason to view the decision that might cause oneself harm more or less ethical depending on if that happens with a low probability or happens with a low presence. 

Price also questions the use of Savage type decision theory. He argues that the decision strategy he calls ``social justice'' is rational but in conflict with the Greaves-Myrvold decision theory. That this strategy should be rational is argued from the rationality of the principle for organizing societies called `social justice.' Again, Price views the descendants as if they are different individuals that exist together in a shared social context, but that is not a correct view of the situation. The isolation of the branches due to decoherence guarantees that the utilities that a rational agent assigns to a branch are independent of the utilities in the other branches unless the `offers' given to the agent cause an artificial correlation.

Albert \cite{albert2010probability} has criticized the work by Greaves and Myrvold by suggesting that he might care more about branches where he is fat because there is more of him there. 
In one way, this is a complaint against their lack of physical reasoning for what value the weights should have. 
The Savage type rationality axioms do not include any such facts about the world, and Greaves and Myrvold were clear that a rule for the values of their branch-weights has to come from some additional arguments. 

The fatness argument also suggests that an agent may have different priorities and wishes when she believes in a branching world than when she believes there is a single outcome. That is indeed possible, but that constitutes no ambiguity for the weights. The weights an agent puts to the different outcomes are independent of the preferences the agent has. The utilities that enter into an agent's decisions are only auxiliary quantities in the analysis of weights or probabilities.

\section{What is Real?\label{sec:real}}

\subsection{The EQM ontology}

The configuration space density (\ref{density}) defined EQM 1 serves an epistemic purpose primarily. It lays out a starting point for the investigations of the world around us. The wavefunction, including all its spin components and eventual gauge indices, describes what exists, but its gauge dependence shows that it also contains something spurious. According to EQM says that particles with distributed positions for their spin components exist\footnote{Allori et al. \cite{allori2011many} have formulated a `primitive ontology,' which is an ontology formulated entirely in ordinary 3-space. Their quantity $m(\vec{x})$ is the sum of all single-particle densities multiplied with respective mass. This mass density does not contain the information that the world consists of individual particles that build up individual items with electromagnetic properties etc. Instead, we have to acknowledge that configuration space is the proper physical space, as it is the space of many particles in 3d-space. If densities are physical, relativity implies that currents are too. In turn, currents imply that spins are physical.}. Relativity implies that currents are equally real. A full discussion of the ontology is postponed to future studies. Anyhow, quantum gravity and related investigations may modify the ontology.

\subsection{Misconceptions}

A misconception about EQM is that the branching creates several copies of the world, which may lead to concerns about energy conservation. However, branching is not copying. Branching is a multi-entanglement creation. To understand the mistake when branching is thought of as copying, consider two electrons that collide, which causes the electrons to become entangled. However, the entanglement does not cause the two electrons to become four electrons. Correspondingly, an observer gets entangled with a detector when recording the measurement value, but this process does not imply copying of the observer into several observers. Instead, the observer becomes distributed into several separate regions in configuration space. The observer's record of the measurement result varies with the position in configuration space. As detector systems are never 100 \% efficient, there will be regions in configuration space where the observer failed to get information about the measured system, bad luck.  

Some criticisms of the derivations of the Born rule are related to the misconception that branching is copying. 
For example, Hemmo and Pitowsky \cite{hemmo2007quantum} contrast the standard quantum mechanics where one alternative becomes ``realized'' with EQM where ``all of them are real''. Price \cite{price2010decisions} wrote about EQM that ``all possible outcomes of quantum measurement are treated as equally real''. With such views, it is no wonder that they come to that the Born rule is a logical impossibility. If we know that something will `really' happen, then it has probability one. 

The EQM description of a single particle is a spatially varying amplitude for each spin component. Its absolute square $\rho_j(\vec{x})$ gives the locations of the particles, which is a distribution. If the particle is in a bound state, it can be probed with forces. The strength of the interactions will reveal the values of $\rho$ in different regions (and spins).  

Is the quantum particle equally real at all positions where the amplitude is non-zero? A quantum particle is not that kind of thing that is localized to a single point. The question presumes something that is not at hand. The same is valid for complex systems as well. They are not localized to a single point in configuration space and spin, but a distributed quantity. The view that all the branches are equally real is a category mistake. It is the whole set of branches with their respective amplitudes that constitute reality.
 Within a branch, that particular branch constitutes the whole reality. What is real depends on the (possible) perspective.

From the pre-measurement perspective, the future observer's experience of the world being one specific branch is an illusion. The views expressed by Hemmo, Pitowsky, and Price that ``all of them are real'' corresponds to viewing mirages as if they are real.

Rovelli \cite{rovelli1996relational} questioned the ideas behind EQM that the observer does not learn a unique value, ``If so, how could have we learned quantum theory?''. However, it has been shown in previous sections that in the overwhelming part of the presence distribution, physicists can deduce the Born rule and confirm quantum mechanics. Note, the pessimist might even claim that if the universe gives single random outcomes, we cannot learn about its nature. If a single outcome were the case, with a small probability, he or she would be right.

\section{Conclusions and Final Remarks \label{section:conclusions}}

The need for assumptions in order to address probabilities have been discussed by Barrett \cite{barrett2017typical}. In the present theory, the assumptions are given by the two postulates and the assumption on the interactions.
The postulate EQM 1 gives a physical foundation to Everett's quantum mechanics. 
It implies that the quantum state belongs to a Hilbert space and an extraordinary simplicity of the postulates. 

The observation of wavefunction structures is a complex process that involves decoherence. 
The previous formulations of decoherence theory were based on the Born rule. 
EQM 1 interprets the quantity appearing in the Born rule $\rho_b$ to give to what extent the system is present at $b$. 
The physical description of measurements assumes that interactions are essentially that of the standard model of particle physics. 
This assumption implies that particle recording detectors only react to the part of the measured state that falls upon the detector. There is no need to explain the measurement process under various hypothetical types of interactions that might not even allow for the construction of detectors. 
Much of the discussions of measurements and the Born rule have gone astray in unnecessarily general and abstract reasonings. The cause for this may be the heritage of classical mechanics and quantum mechanics textbooks, where the mechanics and the interactions are two completely independent entities. 
 
The path to proving the Born rule has been the one proposed by Papineau, which consists of two legs. Firstly, it was shown that statistical inference of $\rho_b$ is possible within the major part of where the system is present. This result implies that EQM is equally well verified as the standard interpretation of quantum mechanics. Secondly, a rational agent will make decisions as if the Born rule quantum probabilities were classical probabilities.  

It is possible to take the quantum state as the full description of the physical world without any additional degrees of freedom or mechanisms that select a single value in a measurement. 
All aspects of the measurement process are fully understood using Everett's interpretation with EQM 1 and EQM 2. 
This fact explains the elusive character of the measurement problem that made Feynman doubt its existence. It proves the suggestion \cite{bohr2004principle} that a selection happens without a cause is correct. As there is no actual selection, there is no cause for it.

%\begin{acknowledgements}
I wish to acknowledge Ben Mottelson, David Wallace, Robert Geroch and Lev Vaidman for stimulating discussions and useful suggestions  
%\end{acknowledgements}

%\bibliography{../Bib_physics_shortTitles}

\begin{thebibliography}{10}
\providecommand{\url}[1]{{#1}}
\providecommand{\urlprefix}{URL }
\expandafter\ifx\csname urlstyle\endcsname\relax
  \providecommand{\doi}[1]{DOI \discretionary{}{}{}#1}\else
  \providecommand{\doi}{DOI \discretionary{}{}{}\begingroup
  \urlstyle{rm}\Url}\fi

\bibitem{everett1957relative}
H.~Everett~III, Rev Mod Phys \textbf{29}(3), 454 (1957)

\bibitem{sep-qm-relational}
F.~Laudisa, C.~Rovelli, in \emph{The Stanford Encyclopedia of Philosophy}, ed.
  by E.N. Zalta, summer 2013 edn. (Metaphysics Research Lab, Stanford
  University, 2013)

\bibitem{rovelli1996relational}
C.~Rovelli, International Journal of Theoretical Physics \textbf{35}(8), 1637
  (1996)

\bibitem{weinberg2015lectures}
S.~Weinberg, \emph{Lectures on quantum mechanics}, 2nd edn. (Cambridge
  University Press, 2015)

\bibitem{hemmo2007quantum}
M.~Hemmo, I.~Pitowsky, Stud Hist Philos Sci B \textbf{38}(2), 333 (2007)

\bibitem{wallace2010prove}
D.~Wallace, in \emph{Many Worlds?: Everett, Quantum Theory, and Reality}, ed.
  by S.~Saunders, J.~Barrett, A.~Kent, D.~Wallace (Oxford University Press,
  2010), pp. 227--263

\bibitem{wallace2012emergent}
D.~Wallace, \emph{The emergent multiverse: Quantum theory according to the
  Everett interpretation} (Oxford University Press, 2012)

\bibitem{deutsch1999quantum}
D.~Deutsch, in \emph{Proc R Soc Lond A}, vol. 455 (The Royal Society, 1999),
  vol. 455, pp. 3129--3137

\bibitem{carroll2014many}
S.M. Carroll, C.T. Sebens, in \emph{Quantum Theory: A Two-Time Success Story}
  (Springer, 2014), pp. 157--169

\bibitem{sebens2016self}
C.T. Sebens, S.M. Carroll, Brit J Philos Sci \textbf{69}(1), 25 (2016)

\bibitem{pittphilsci14389}
K.J. McQueen, L.~Vaidman.
\newblock In defence of the self-location uncertainty account of probability in
  the many-worlds interpretation.
\newblock Electronic (2018).
\newblock \urlprefix\url{http://philsci-archive.pitt.edu/14389/}

\bibitem{pittphilsci8558}
L.~Vaidman, \emph{Probability in Physics} (Springer, Berlin, Heidelberg, 2012),
  chap. Probability in the Many-Worlds Interpretation of Quantum Mechanics, pp.
  299--311

\bibitem{kent2015does}
A.~Kent, Found Phys \textbf{45}(2), 211 (2015)

\bibitem{KentAgainstMWI}
A.~Kent, Int J Mod Phys A \textbf{5}, 1745 (1990)

\bibitem{maudlin2014critical}
T.~Maudlin, No{\^u}s \textbf{48}(4), 794 (2014)

\bibitem{joos2000elements}
E.~Joos, in \emph{Decoherence: Theoretical, experimental, and conceptual
  problems} (Springer, 2000), pp. 1--17

\bibitem{BornRuleEnvariance}
W.H. Zurek, Phys Rev A \textbf{71}, 052105 (2005)

\bibitem{QDarwinPhysToday}
W.H. Zurek, Phys Today \textbf{67}(10), 44 (2014)

\bibitem{barrett2017typical}
J.A. Barrett, Stud Hist Philos Sci B \textbf{58}, 31 (2017)

\bibitem{albert2010probability}
D.~Albert, in \emph{Many Worlds?: Everett, Quantum Theory, and Reality}, ed. by
  S.~Saunders, J.~Barrett, A.~Kent, D.~Wallace (Oxford University Press, 2010),
  pp. 355--368

\bibitem{kent2010one}
A.~Kent, in \emph{Many Worlds?: Everett, Quantum Theory, and Reality}, ed. by
  S.~Saunders, J.~Barrett, A.~Kent, D.~Wallace (Oxford University Press, 2010),
  pp. 307--354

\bibitem{vaidman1998schizophrenic}
L.~Vaidman, Int Stud Philos Sci \textbf{12}(3), 245 (1998)

\bibitem{greaves2004understanding}
H.~Greaves, Stud Hist Philos Sci B \textbf{35}(3), 423 (2004)

\bibitem{greaves2007everettian}
H.~Greaves, Stud Hist Philos Sci B \textbf{38}(1), 120 (2007)

\bibitem{SchrodingerIV}
E.~Schr{\"o}dinger, Annalen der Physik \textbf{81}, 109 (1926)

\bibitem{negele1970structure}
J.W. Negele, Phys Rev C \textbf{1}(4), 1260 (1970)

\bibitem{joos2013decoherence}
E.~Joos, H.D. Zeh, C.~Kiefer, D.J. Giulini, J.~Kupsch, I.O. Stamatescu,
  \emph{Decoherence and the appearance of a classical world in quantum theory}
  (Springer Science \& Business Media, 2013)

\bibitem{nielsen2010quantum}
M.A. Nielsen, I.L. Chuang, \emph{Quantum computation and quantum information}
  (Cambridge university press, 2010)

\bibitem{GleasonMeasureOnHS}
A.M. Gleason, Journal of Mathematics and Mechanics \textbf{6}, 885 (1957)

\bibitem{Geroch1984}
R.~Geroch, No{\^u}s \textbf{18}(4), pp. 617 (1984)

\bibitem{vaidman2016all}
L.~Vaidman, in \emph{J Phys Conf Ser}, vol. 701 (IOP Publishing, 2016), vol.
  701, p. 012020

\bibitem{zeh1970interpretation}
H.D. Zeh, Found Phys \textbf{1}(1), 69 (1970)

\bibitem{joos1985emergence}
E.~Joos, H.D. Zeh, Z Physik B \textbf{59}(2), 223 (1985)

\bibitem{baker2007measurement}
D.J. Baker, Studies In History and Philosophy of Science Part B: Studies In
  History and Philosophy of Modern Physics \textbf{38}(1), 153 (2007)

\bibitem{dawid2015many}
R.~Dawid, K.P. Th{\'e}bault, Synthese \textbf{192}(5), 1559 (2015)

\bibitem{zurek2013wave}
W.H. Zurek, Phys Rev A \textbf{87}(5), 052111 (2013)

\bibitem{feller1968introduction}
W.~Feller, \emph{An Introduction to Probability Theory and Its Applications:
  Volume 1}, 3rd edn. (J. Wiley \& sons, 1968)

\bibitem{SquiresOnProof}
E.J. Squires, Phys Lett A \textbf{145}(2,3), 67 (1990)

\bibitem{papineau1996many}
D.~Papineau, The British Journal for the Philosophy of Science \textbf{47}(2),
  233 (1996)

\bibitem{greaves2010everett}
H.~Greaves, W.~Myrvold, in \emph{Many Worlds?: Everett, Quantum Theory, and
  Reality}, ed. by S.~Saunders, J.~Barrett, A.~Kent, D.~Wallace (Oxford
  University Press, 2010), pp. 264--307

\bibitem{feller2008introduction}
W.~Feller, \emph{An introduction to probability theory and its applications},
  vol.~2 (John Wiley \& Sons, 2008)

\bibitem{appleby2005facts}
D.~Appleby, Found Phys \textbf{35}(4), 627 (2005)

\bibitem{appleby2005probabilities}
D.~Appleby, Opt Spectrosc+ \textbf{99}(3), 447 (2005)

\bibitem{caves2005properties}
C.M. Caves, R.~Schack, Annals of Physics \textbf{315}(1), 123 (2005)

\bibitem{hartle1968quantum}
J.B. Hartle, Am J Phys \textbf{36}(8), 704 (1968)

\bibitem{gutmann1995using}
S.~Gutmann, Phys Rev A \textbf{52}(5), 3560 (1995)

\bibitem{lewis1980subjectivist}
D.~Lewis, in \emph{Ifs} (Springer, 1980), pp. 267--297

\bibitem{de2017theory}
B.~De~Finetti, \emph{Theory of probability: a critical introductory treatment}
  (John Wiley \& Sons, 2017)

\bibitem{savage1972foundations}
L.J. Savage, \emph{The foundations of statistics} (Courier Corporation, 1972)

\bibitem{wallace2002quantum}
D.~Wallace, arXiv preprint quant-ph/0211104  (2002)

\bibitem{wallace2007quantum}
D.~Wallace, Stud Hist Philos Sci B \textbf{38}(2), 311 (2007)

\bibitem{greaves2007probability}
H.~Greaves, Philosophy Compass \textbf{2}(1), 109 (2007)

\bibitem{saunders2008branching}
S.~Saunders, D.~Wallace, The British Journal for the Philosophy of Science
  \textbf{59}(3), 293 (2008)

\bibitem{saunders2010chance}
S.~Saunders, in \emph{Many Worlds?: Everett, Quantum Theory, and Reality}, ed.
  by S.~Saunders, J.~Barrett, A.~Kent, D.~Wallace (Oxford University Press,
  2010), pp. 181--205

\bibitem{duhamel2011guildenstern}
V.~Duhamel, P.~Raymond-Robichaud, arXiv preprint arXiv:1111.2563  (2011)

\bibitem{price2010decisions}
H.~Price, in \emph{Many Worlds?: Everett, Quantum Theory, and Reality}, ed. by
  S.~Saunders, J.~Barrett, A.~Kent, D.~Wallace (Oxford University Press, 2010),
  pp. 369--391

\bibitem{allori2011many}
V.~Allori, S.~Goldstein, R.~Tumulka, N.~Zangh{\`\i}, The British Journal for
  the Philosophy of Science \textbf{62}(1), 1 (2011)

\bibitem{bohr2004principle}
A.~Bohr, B.R. Mottelson, O.~Ulfbeck, Found Phys \textbf{34}(3), 405 (2004)

\end{thebibliography}

\end{document}